%% file: ISIT2012-songnam.tex
\documentclass[conference]{IEEEtran}
\usepackage{cite}
\usepackage{graphicx,color,epsfig,rotating}
\usepackage{amsfonts,amsmath,amssymb}
\usepackage{algorithm,algorithmic}
\usepackage{subfigure}
\input{macros}

\newtheorem{definition}{Definition}

\newtheorem{remark}{Remark}

\begin{document}

\title{Reverse Compute and Forward: \\ A Low-Complexity Architecture for Downlink Distributed Antenna Systems}

\author{
\IEEEauthorblockN{Song-Nam Hong}
\IEEEauthorblockA{
University of Southern California \\
Los Angeles, CA 90089}
\and
\IEEEauthorblockN{Giuseppe Caire}
\IEEEauthorblockA{
University of Southern California \\
Los Angeles, CA 90089}
}


\maketitle

\begin{abstract}
We consider a distributed antenna system where $L$ antenna terminals (ATs) are connected to a Central Processor (CP)
via digital error-free links of finite capacity $R_0$, and serve $L$ user terminals (UTs).
This system model has been widely investigated both for the uplink
and the downlink, which are instances of the general multiple-access relay and broadcast relay networks.
In this work we focus on the downlink, and propose a novel downlink precoding scheme nicknamed ``Reverse Quantized Compute and Forward'' (RQCoF). For this scheme we obtain achievable rates and compare with the state of the art available in the literature.
We also provide simulation results for a realistic network with fading and pathloss with $K > L$ UTs,
and show that channel-based user selection produces large benefits and essentially
removes the problem of rank deficiency in the system matrix.\footnote{This research was supported in part by the KCC (Korea Communications Commission), Korea, under the R\&D program supervised by the KCA (Korea Communications Agency) (KCA-2011-11921-04001).}
\end{abstract}


\section{System and problem definition} \label{sec:intro}

We consider a distributed antenna system (DAS) with $K$ user terminals (UTs) and  $L$ ``antenna terminals" (ATs).
All UTs and ATs have a single antenna each.
The ATs are connected with a central processor (CP) via wired links of fixed rate $R_{0}$.
We study the downlink scenario, where the CP wishes to deliver independent messages to the UTs.
This is a simple instance of a broadcast relay network, where the ATs operate as relays.
In this work we focus on the symmetric rate, i.e., all messages have the same rate
and assume that the CP and all UTs have perfect channel state information (more general results are provided in \cite{songnam2011DAS}).
If $R_{0} \rightarrow \infty$, the problem reduces to the well-known vector Gaussian broadcast channel, the capacity region of which is achieved by
Dirty Paper Coding (DPC). However, for fixed finite $R_{0}$, DPC and other widely considered linear precoding schemes cannot be applied
in a straightforward manner. A simple DAS system, the so-called Soft-Handoff model, was investigated in \cite{simeone2009downlink},
by introducing a ``compressed" version of DPC (CDPC), where the CP performs joint DPC under per-antenna power constraint and then sends
the compressed (or quantized) codewords to the corresponding ATs via the wired links. While this scheme is expected to be near-optimal
for very large $R_{0}$, it is generally suboptimal at finite (possibly small) $R_{0}$. Also, DPC is notoriously difficult to be implemented in practice,
due to the nested lattice coding construction and lattice quantization steps involved (See for example \cite{Erez2005,bennatan2006superposition}).

Motivated by Compute-and-Forward (CoF) \cite{nazer2011compute} (or quantized compute-and-forward (QCoF) \cite{songnam2011compute}),
we propose a novel coding strategy named Reverse QCoF (RQCoF) for the DAS downlink with finite backhaul link capacity $R_{0}$.
In QCoF and RQCoF the coding block length $n$ can be arbitrarily large but
the shaping block length is restricted to 1 (scalar quantization \cite{songnam2011compute}).
However, we would like to point out that the same approach can be straightforwardly applied to
CoF based schemes, where also the shaping dimension becomes large (in this case, we would refer to the scheme as Reverse CoF (RCoF)).

\subsection{Overview of QCoF}\label{sec:QCoF}

Let $\ZZ_{p} = \ZZ \mod p\ZZ$ denote the finite field of size $p$, with $p$ a prime number, $\oplus$ denote addition
over $\ZZ_p$, and $g : \ZZ_p \rightarrow \RR$ be the natural mapping of the elements of $\ZZ_p$ onto $\{0,1,...,p-1\} \subset \RR$.
For a lattice $\Lambda$, let $Q_{\Lambda}(\xv) = {\rm argmin}_{\lambdav \in \Lambda} \{  \| \xv - \lambdav\| \}$ denote the associated lattice quantizer,
$\Vc = \{ \xv \in \RR^n : Q_\Lambda(\xv) = \zerov\}$ the Voronoi region and define
$[\xv ] \mod \Lambda =  \xv - Q_{\Lambda}(\xv)$.
For $\kappa \in \RR$, consider the two nested one-dimensional lattices
$\Lambda_s  =  \{x = \kappa p z: z \in \ZZ\}$ and $\Lambda_c   =  \{x = \kappa z: z \in \ZZ\}$,
and define the {\em constellation set} $\Sc \triangleq \Lambda_c \cap \Vc_s$,
where $\Vc_s$ is the {\em Voronoi region} of $\Lambda_s$, i.e., the interval $[-\kappa p/2, \kappa p/2)$.
The {\em modulation mapping} $m : \ZZ_p \rightarrow \Sc$ is defined by
$v = m(u) \triangleq [\kappa g(u)] \mod \Lambda_s$. The inverse function $m^{-1}(\cdot)$ is referred to as the {\em demodulation mapping},
and it is given by  $u = m^{-1}(v) \triangleq g^{-1}( [ v/\kappa ]  \mod p \ZZ)$ with $v \in \Sc$.

Consider the (real-valued) $L$-user Gaussian multiple access channel  with inputs $\{x_{\ell,i}: i=1,...,n\}$ for $\ell=1,...,L$, output $\{y_{i}:i=1,...,n\}$
and coefficients $\hv=(h_{1},...,h_{L})^{\transp} \in \RR^{L}$, defined by
\begin{equation}
y_{i} = \sum_{i=1}^{L} h_{\ell}x_{\ell,i} + z_{i}, \;\; \hbox{ for } \;\; i=1,\ldots,n,
\end{equation}
where the $z_{i}$'s are i.i.d. $\sim \Nc(0,1)$. All users encode their information messages $\{\wv_{\ell} \in \ZZ_{p}^{k}: \ell=1,\ldots,L\}$
using the same linear code $\Cc$ over $\ZZ_{p}$ (i.e., denoting information sequences and codewords by row vectors,
we have $\cv_{\ell}= \wv_{\ell} \Gm$ where $\Gm$ is a generator matrix for $\Cc$),
and produce their channel inputs according to
\begin{equation}\label{eq:channelinput}
x_{\ell,i} = [m(c_{\ell,i}) + d_{\ell,i}] \mod \Lambda_{s}, \;\;  i=1,\ldots,n,
\end{equation}
where $c_{\ell,i}$ is the $i$-th symbol of $\cv_\ell$ and $d_{\ell,i}$'s are i.i.d. dithering symbols
$\sim \mbox{Uniform}(\Vc_s)$, known at the receiver.
The channel inputs $x_{\ell,i}$ are uniformly distributed over $\Vc_s$ and have second moment
$\SNR \triangleq \EE[|x_{\ell,i}|^{2}] = \kappa^2 p^2 / 12$.
The receiver's goal is to recover a linear combination $\cv = \bigoplus q_{\ell}\cv_{\ell}$ of the transmitted users' codewords,
for some coefficients $q_{\ell} \in \ZZ_{p}$.
For this purpose, the receiver selects the \emph{integer coefficients vector} $\av= (a_{1},...,a_{L})^{\transp} \in \ZZ^{L}$
and produces the sequence of quantized observations
\begin{equation}\label{eq:demo}
u_{i} =   m^{-1} \left ( \left [Q_{\Lambda_c} \left (\alpha y_{i} - \av^{\transp}\dv_{i} \right ) \right ] \mod \Lambda_s \right ),
\end{equation}
for $i=1, \ldots ,n$. It easy to show \cite{songnam2011compute} that (\ref{eq:demo}) is equivalent to
\begin{equation}\label{eq:model}
u_{i} = \Big( \bigoplus_{\ell=1}^{L} q_{\ell} c_{\ell,i} \Big) \oplus \tilde{z_{i}},
\end{equation}
with $q_{\ell} = g^{-1}([a_{\ell}] \mod p\ZZ)$. Here, $\tilde{z_{i}} = m^{-1}([Q_{\Lambda_c}(\varepsilon)] \mod \Lambda_{s})$ where $\varepsilon$ denotes the effective noise, capturing a Gaussian additive noise and \emph{non-integer penalty},
and its variance \cite{songnam2011compute} is
\begin{equation}\label{eq:eff-variance}
\sigma_{\varepsilon}^2 = \av^{\transp}(\SNR^{-1}\Id + \hv\hv^{\transp})^{-1}\av.
\end{equation}
By \cite[Th. 1]{songnam2011compute}, the achievable computation rate of QCoF is given by
\begin{equation}\label{QCoFrate}
R_{\texttt{\textrm{\tiny QCoF}}} = \log{p} - H(\tilde{z}).
\end{equation}
Also, by \cite[Th. 4]{nazer2011compute}, the achievable computation rate of CoF is given by
\begin{equation}\label{CoFrate}
R_{\texttt{\textrm{\tiny CoF}}}(\sigma_{\varepsilon}^2) = \frac{1}{2}\log(\SNR/\sigma_{\varepsilon}^2).
\end{equation}
We showed in \cite{songnam2011compute} that, for fixed large $\SNR \gg 1$ and sufficiently large $p$ (e.g., $p \geq 251$), the
(\ref{QCoFrate}) and (\ref{CoFrate}) differ approximately by the shaping gain, i.e., $\approx 0.25$ bits per real dimension.

\begin{remark}
In order to achieve the CoF rate, $p$ must grow to infinity in the lattice construction and the rank of the system matrix $\Qm$
is the same as the rank of $\Am$ over $\RR$, by \cite[Th. 11]{nazer2011compute}.
\end{remark}

\section{Reverse Quantized Compute-and-Forward}\label{sec:RQCoF}

The main idea is that each UT decodes a linear combination (over the finite field) of the messages sent by the ATs using QCoF.
In short, we exchange the role of the ATs and UTs and use QCoF in the reverse direction.
However, decoding linear combination of the information messages is useful only when these combinations can be shared such
that the individual messages can be recovered, provided that the resulting system of linear equations is invertible over $\ZZ_{p}$.
Since the UTs do not cooperate, sharing the decoded linear combinations is impossible in the downlink.
Nevertheless, thanks to algebraic structure of QCoF (or CoF), the messages from the ATs can be the precoded versions of the original information
messages and hence, using an appropriate invertible precoding over $\ZZ_{p}$ at the CP, the effect of the linear combination can be undone
at the transmitter, so that every UT obtains just its own desired message.
We present coding strategies considered in this work, assuming the $K = L$ and (real-valued) channel matrix
$\Hm \in \RR^{L \times L}$.
Let $\Qm$ denote the system matrix whose elements in the $\ell$-th row, denoted by $\qv_{\ell}^{\transp}=(q_{\ell,1},...,q_{\ell,L})$,
indicate the coefficients of the linear combination decoded at the $\ell$-th UT as given in (\ref{eq:model}).
For the time being we assume that these matrices are full rank over $\ZZ_{p}$, although they may be rank deficient since each UT chooses
its own linear combination coefficients independently of the other nodes.
The case of rank deficiency will be handled later.
Let $\tilde{z}_{\ell}$ be the discrete additive noise (over $\ZZ_{p}$) at the $\ell$-th UT.
The detailed description of ``reverse" QCoF (RQCoF) is as follows.
\begin{itemize}
\item
For the given $\Qm$, the CP precodes the user information messages $\{\wv_{\ell} \in \ZZ_{p}^{k}:\ell=1,...,L\}$ using the inverse system
matrix $\Qm^{-1}$. The precoded $L$-dimensional vectors of information symbols to be transmitted by the ATs are given by
\begin{equation} \label{precoded-messages}
(\mu_{1,i},...,\mu_{L,i})^{\transp} = \Qm^{-1}(w_{1,i},...,w_{L,i})^{\transp}, \;\; \mbox{for} \;\; i=1,\ldots,k.
\end{equation}
\item The CP forwards each block $\muv_{\ell}=(\mu_{\ell,1},...,\mu_{\ell,k})$ to the $\ell$-th AT, during $n$ time slots, corresponding
to the duration of a codeword sent on the wireless channel. Therefore, we have the rate constraint $(k/n) \log{p} \leq R_{0}$.
\item After receiving $k$ symbols, the $\ell$-th AT locally encodes its information symbols $\muv_{\ell}$ using the same linear
code $\Cc$ over $\ZZ_{p}$ (i.e., $\cv_{\ell} = \muv_{\ell}\Gm$), and produces its channel input according to
\begin{equation}\label{eq:channelinput2}
x_{\ell,i} = [m(c_{\ell,i}) + d_{\ell,i}] \mod \Lambda_{s}, \;\; \hbox{ for } \;\; i=1,\ldots,n.
\end{equation}
\item By \cite[Th. 1]{songnam2011compute}, the $\ell$-th UT can recover a noiseless linear combination of \emph{ATs' information symbols} if $R \leq \log{p} - \max_{\ell}\{H(\tilde{z}_{\ell})\}$. This is given by
\begin{eqnarray}
\qv_{\ell}^{\transp}(\mu_{1,i},...,\mu_{L,i})^{\transp} &=& \qv_{\ell}^{\transp}\Qm^{-1}(w_{1,i},...,w_{L,i})^{\transp} \nonumber \\
&=& w_{\ell,i}, \hbox{ for }\;\; i=1,\ldots,k. \nonumber \\
& & \label{eq:undone2}
\end{eqnarray}
Hence, the $\ell$-th UT can successfully recover its desired message.
\end{itemize}
The following rate is achievable by RQCoF:
\begin{equation}\label{eq:rateRQCoF}
R_{\texttt{\textrm{\tiny RQCoF}}} = \min\{R_{0},\log{p} - \max_{\ell}\{H(\tilde{z}_{\ell})\}.
\end{equation}
Similarly, from (\ref{CoFrate}), we can get an achievable rate per user of RCoF
\begin{equation}\label{eq:rateRCoF}
R_{\texttt{\textrm{\tiny RCoF}}} = \min\{R_{0},\min_{\ell}\{R_{\texttt{\textrm{\tiny CoF}}}(\sigma_{\varepsilon_{\ell}}^2)\}\}.
\end{equation} Finally, the achievable rate of RQCoF (or RCoF) is maximized by minimizing the variance of effective noise in (\ref{eq:eff-variance}) with respect to $\Am$ subject to the system matrix $\Qm$ is full rank over $\ZZ_{p}$. This problem was solved in \cite{songnam2011compute} using  the LLL algorithm\cite{lenstra1982factoring}, possibly followed by Phost or Schnorr-Euchner enumeration (See  \cite{schnorr1994lattice}) of the non-zero lattice points in a sphere centered at the origin, with radius equal to the shortest vector found by LLL.

\section{Compressed Integer-Forcing Beamforming}

In short, the idea underlying RQCoF is that each UT converts  its own downlink channel into a discrete additive-noise multiple access  channel
over $\ZZ_p$. Since each UT is interested only in its own message, the CP can precode the messages using zero-forcing linear precoding
over $\ZZ_p$, at no transmit power additional cost (unlike linear zero-forcing over $\RR$).
It is known that the performance of CoF (and therefore QCoF) is quite sensitive to the channel coefficients,
due to the non-integer penalty, since the channel coefficients are not exactly matched to the integer coefficients of linear
combinations \cite{nazer2011compute,songnam2011compute}.
The same problem arises in RQCoF (or RCoF), due to their formal equivalence.
In \cite{Zhan2011IFLR}, it was shown that integer-forcing linear receiver (IFLR) can eliminate this penalty by forcing the effective
channel matrix to be integer. Here, we propose a new beamforming strategy named Integer-Forcing Beamforming (IFBF),
that produces a similar effect for the downlink.

We present the IFBF idea assuming $R_{0} = \infty$, as the dual scheme of IFLR, and consider finite $R_0$ later.
In IFBF, the beamforming vectors $\Wm = [\wv_{1},...,\wv_{L}]$ are chosen such that the effective channel matrix $\tilde{\Hm} = \Hm\Wm$ is integer-valued.
Then, the channel matrix is inverted over $\ZZ_q$ by using RQCoF, as previously presented. In this case, since
$\tilde{\Hm} \in \ZZ^{L \times L}$, RQCoF  does not suffer from the non-integer penalty.
Further, we extend IFBF to the case of finite $R_{0}$ by using quantization, as in done in \cite{simeone2009downlink},
where CP forwards the quantized sequences to the ATs for which the quantization noise is determined from standard rate-distortion theory bounds.
It is assumed that $\Hm \in \RR^{L \times L}$ is full rank and the detailed description of IFBF is as follows.
For a given $\Am \in \ZZ^{L \times L}$ (optimized later),  the CP uses the beamforming matrix $\Wm = \Hm^{-1}\Am$
and the system matrix $\Qm = [\Am] \mbox{ mod } p\ZZ$ as in Section \ref{sec:QCoF}.

Assuming that $\Qm$ is full rank over $\ZZ_{p}$,  the CP produces the downlink streams $\xv_\ell=\{x_{\ell,i}:i=1,\ldots,n\}$, for $\ell = 1,\ldots, L$ as follows.
\begin{itemize}
\item
The CP precodes the user information messages $\{\wv_{\ell} \in \ZZ_{p}^{k}:\ell=1,...,L\}$ using the inverse system
matrix $\Qm^{-1}$:
\begin{equation} \label{precoded-messages2}
(\mu_{1,i},...,\mu_{L,i})^{\transp} = \Qm^{-1}(w_{1,i},...,w_{L,i})^{\transp},
\end{equation} $\;\; \mbox{for} \;\; i=1,\ldots,k.$
\item The CP encodes the precoded information messages using the same linear code $\Cc$ over $\ZZ_{p}$  (i.e., $\cv_{\ell} = \muv_{\ell}\Gm$) and produces the downlink stream according to
\begin{equation}\label{eq:channelinput3}
x_{\ell,i} = [m(c_{\ell,i}) + d_{\ell,i}] \mod \Lambda_{s}, \;\; \hbox{ for } \;\; i=1,\ldots,n.
\end{equation}
\end{itemize}
Using the predefined $\Wm$, the CP produces the precoded channel inputs $\{v_{\ell,i}:i=1,\ldots,n\}$ using
\[  (v_{1,i}, \ldots, v_{L,i})^\transp = \Wm (x_{1,i}, \ldots, x_{L,i})^\transp, \;\; \mbox{for} \;\; i = 1,\ldots, n, \]
and forwards them to the ATs via the wired links.
Consistently with our system definition, we impose a per-antenna power constraint equal to $\SNR$ (with suitable normalization).
Hence, the second moment of $x_{\ell,i}$ is determined as
\begin{equation}
\EE[|x_{\ell,i}|^{2}] = \SNR/\max_{\ell}\{\|\Hm^{-1}\av_{\ell}\|^2\},
\end{equation}
which guarantees that the power of the signal transmitted from the $\ell$-th AT has the required power
$\EE[|v_{\ell,i}|^2] = \SNR$.  The received signal at the $\ell$-th UT is given by
\begin{equation}
y_{\ell,i} = \av_{\ell}^{\transp} x_{\ell,i} + z_{\ell,i}, \;\; \mbox{for} \;\; i = 1, \ldots, n.
\end{equation}
Notice that thanks to the IFBF the non-integer penalty is equal to zero.
So, every UT can recover its desired messages by decoding the linear combination of ATs' messages with integer
coefficients $\av_{\ell}$ as shown in (\ref{eq:undone2}).
Finally, the achievable rate of IFBF with RQCoF can be obtained by numerically computing the entropy of discrete additive
noise over $\ZZ_{p}$ corresponding to effective noise $\varepsilon_{\ell} \sim \Nc(0,\max_{\ell}\{||\Hm^{-1}\av_{\ell}||^2\})$ where the
impact of power constraint is included in the effective noise. The following rate is achievable by IFBF with RQCoF:
\begin{equation}\label{rate:IFBF}
R_{\texttt{\textrm{\tiny IFBF}}}=\log{p} - \max_{\ell}\{H(\tilde{z}_{\ell})\}
\end{equation}for any full-rank matrix $\Qm$, where $\tilde{z}_{\ell}=m^{-1}([Q_{\Lambda_c}(\varepsilon_{\ell})] \mod \Lambda_{s})$.
From (\ref{CoFrate}), the following rate is achievable by IFBF with RCoF:
\begin{equation}\label{rate:IFBF}
R_{\texttt{\textrm{\tiny IFBF}}}=\frac{1}{2} \log(\SNR/\max_{\ell}\{||\Hm^{-1}\av_{\ell}||^2\})
\end{equation}for any full-rank integer matrix $\Am$.

For the case of finite $R_{0}$, we propose a ``compressed" IFBF (CIFBF) where
the CP forwards the quantized channel inputs $\hat{v}_{\ell,i} = v_{\ell,i} + \hat{z}_{\ell,i}$ for $i=1,\ldots,n$, to the $\ell$-th AT,
where $\{\hat{z}_{\ell,i}:i=1,\ldots,n\}$ denotes the quantization noise sequence, with variance (quantization mean-square error) equal to
$\sigma_{\hat{z}}^{2}$.  From the standard rate-distortion theory, the CP can forward the $\{\hat{v}_{\ell,i}:i=1,...,n\}$ to the $\ell$-th AT if
\begin{equation}\label{eq:cond}
R_{0} \geq I(v_{\ell};\hat{v}_{\ell}),
\end{equation}
where the index $i$ is omitted for brevity.
Using the well-known maximum entropy argument on (\ref{eq:cond}) we have the bound
\begin{equation}\label{eq:bound}
I(v_{\ell};\hat{v}_{\ell}) \leq \frac{1}{2}\log(\SNR/\sigma_{\hat{z}}^{2}).
\end{equation}
From (\ref{eq:cond}) and (\ref{eq:bound}), we obtain $\sigma_{\hat{z}}^2 = \SNR/2^{2R_{0}}$ and $\EE[|v_{\ell,i}|^2] = \SNR/(1+1/(2^{2R_{0}}-1))$,
due to the power constraint.  Accordingly, we have
\begin{equation}
\EE[|x_{\ell,i}|^2]=\SNR/\max_{\ell}\{||\Hm^{-1}\av_{\ell}||^2\}(1+1/(2^{2R_{0}}-1)).
\end{equation}
Also, the effective noise at the $\ell$-th UT is given by
\begin{equation}
\varepsilon_{\ell,i} = z_{\ell,i} + \sum_{k=1}^{L}h_{\ell,k}\hat{z}_{k,i},
\end{equation}
where the second term captures the impact of quantization noise and its variance is
\begin{equation}
\sigma_{\varepsilon_{\ell}}^2 = 1+||\hv_{\ell}||^2\SNR/2^{2R_{0}}.
\end{equation} Finally, the achievable rate of CIFBF with RQCoF can be obtained numerically computing the entropy of discrete additive noise over $\ZZ_{p}$ corresponding to the  effective noise $\varepsilon_{\ell}' \sim \Nc(0,\sigma_{\varepsilon_{\ell}'}^{2})$ where the impact of power constraint and quantization noise are included in the effective noise:
\begin{equation}
\sigma_{\varepsilon_{\ell}'}^{2} = \max_{\ell}\{||\Hm^{-1}\av_{\ell}||^2\}(1+(1+||\hv_{\ell}||^2\SNR)/(2^{2R_{0}}-1)).
\end{equation}  The following rate is achievable by CIFBF with RQCoF:
\begin{equation}\label{rate:IFBF}
R_{\texttt{\textrm{\tiny CIFBF}}}=\log{p} - \max_{\ell}\{H(\tilde{z}_{\ell})\}
\end{equation}for any full-rank matrix $\Qm$, where $\tilde{z}_{\ell}=m^{-1}([Q_{\Lambda_c}(\varepsilon_{\ell}')] \mod \Lambda_{s})$. From the (\ref{CoFrate}), the following rate is achievable by CIFBF with RCoF:
\begin{equation*}
R_{\texttt{\textrm{\tiny CIFBF}}}= R_{\texttt{\textrm{\tiny IFBF}}} - \frac{1}{2}\max_{\ell}\{\log(1+(1+||\hv_{\ell}||^2\SNR)/(2^{2R_{0}}-1))\}
\end{equation*}
for any full-rank integer matrix $\Am$.
Finally, the achievable rate is maximized by minimizing the $\max_{\ell}\{||\Hm^{-1}\av_{\ell}||^2\}$ subject to full-rank constraint. This problem can be thought of as finding  the $L$ linearly independent ``shortest lattice points" of the $L$-dimensional lattice generated by $\Hm^{-1}$. This can be efficiently obtained using the LLL algorithm \cite{lenstra1982factoring}. Specifically, for a given lattice $\Lambda$ defined by $\Lambda=\{x=\Hm^{-1}\zv: \zv \in \ZZ^{L}\}$, a reduced basis of lattice is obtained through a unimodular matrix $\Um$ such that $\Lambda=\{x=\Hm^{-1}\Um\zv: \zv \in \ZZ^{L}\}$. Let $\Fm = \Hm^{-1}\Um$ generates the same lattice but has ``reduced" columns, i.e., the columns of $\Fm$ have small $2$-norm. The solution of the original problem can be chosen as $\av_{\ell} =\uv_{\ell}$ where $\uv_{\ell}$ denotes the $\ell$-th column of $\Um$. While finding the optimal reduced basis for a lattice (e.g., finding the optimal $\Um$) is an NP-hard problem, the LLL algorithm finds a good reduced basis with low-complexity \cite{lenstra1982factoring}.

\begin{remark} In CIFBF, relays (i.e., the distributed antenna elements) have very low-complexity and are oblivious
to codebooks since they just forward the received signals from CP, not requiring modulation and  encoding.
\end{remark}

\begin{remark} In terms of performance, it is worthwhile understand the impact of \emph{non-integer penalty} and \emph{quantization noise}
depending on parameters $R_{0}$, $\SNR$, and so on.
As $R_{0} \rightarrow \infty$, the effect of quantization noise vanishes and thus, CIFBF would be better than RCoF.
However, when $R_{0}$ is small, RCoF without beamforming may perform better than CIFBF since quantization noise would
be severe in this case.  A numerical result in a particular case is  provided in Fig.~\ref{simulation_downlink}.
\end{remark}

\section{Scheduling and Numerical Results}

\begin{figure}
\centerline{\includegraphics[width=9cm]{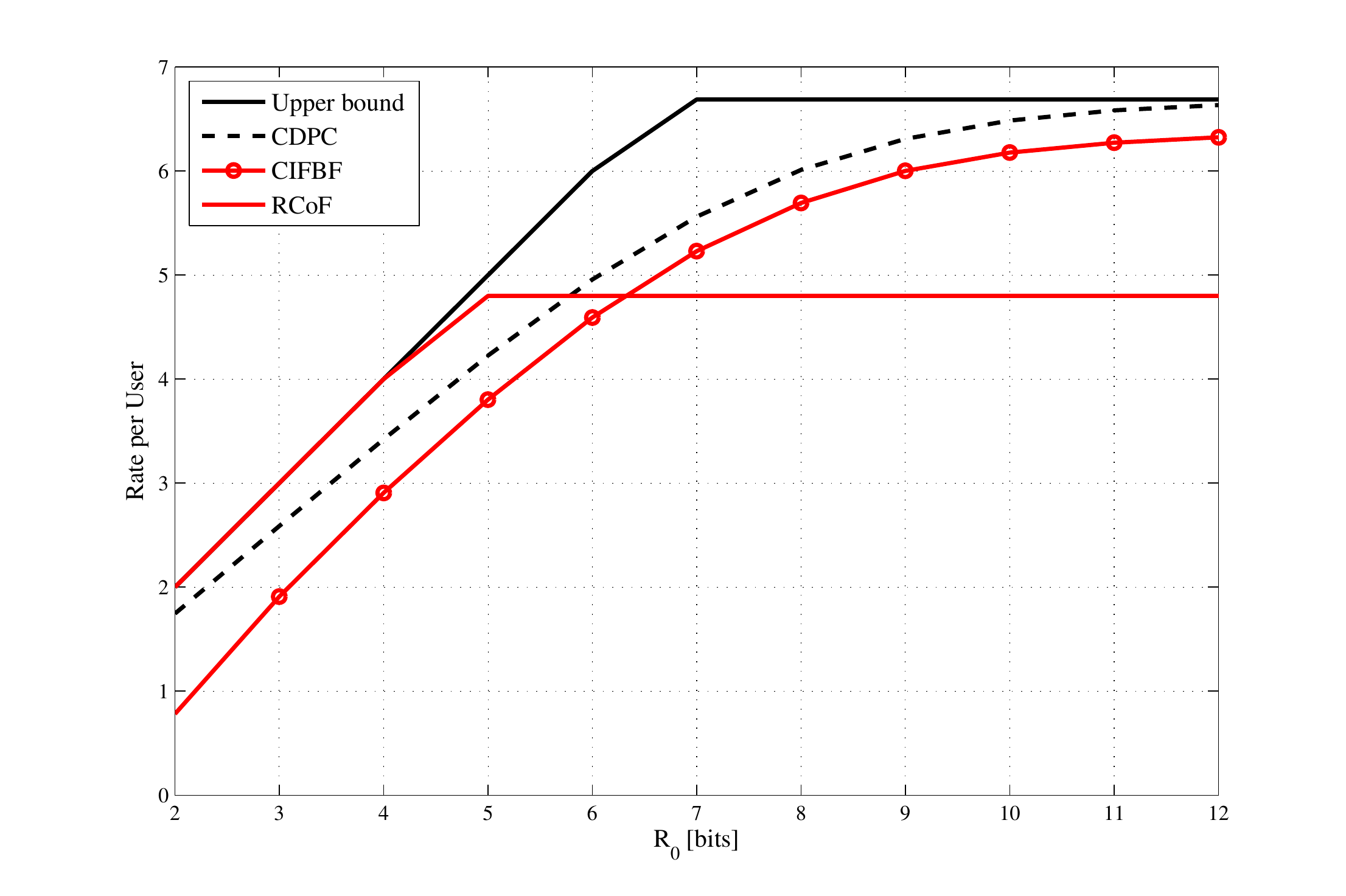}}
\caption{$\SNR = 20$dB. Achievable rates per user as a function of finite capacity $R_{0}$, for inter-cell interference
$\gamma \sim \mbox{Uniform}(0.5,1)$.}
\label{simulation_downlink}
\end{figure}

\begin{figure}
\centerline{\includegraphics[width=9cm]{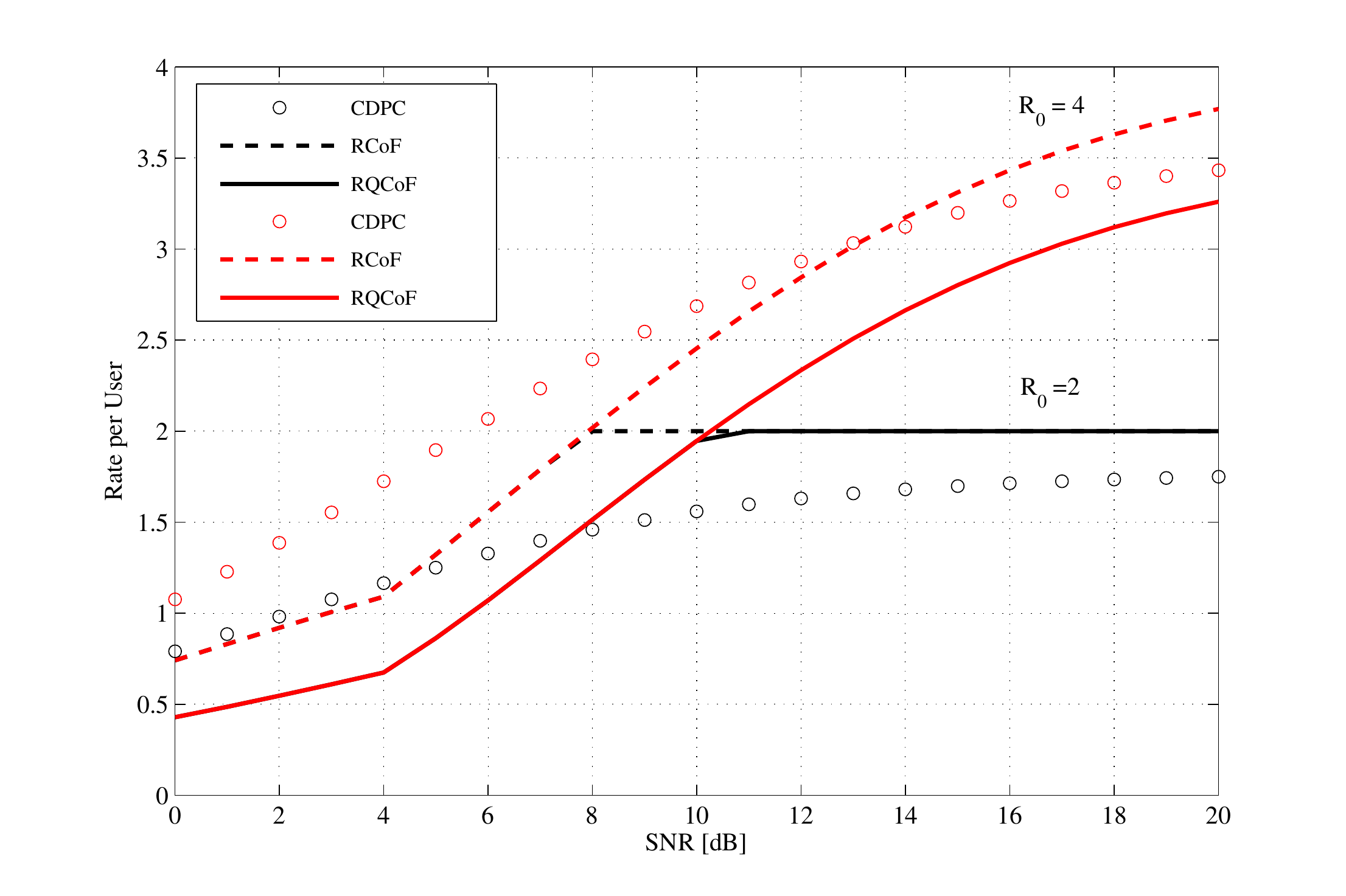}}
\caption{Achievable rates per user as a function of SNRs,
for finite capacity $R_{0}=2$ or $4$ bits, inter-cell interference level $\gamma = 0.7$, and $p=251$ for RQCoF.}
\label{simulation_DL}
\end{figure}

For the sake of comparison with CDPC we consider the same Soft-Handoff model of \cite{simeone2009downlink},
with $L$ ATs and $L$ UTs for which the received signal at the $\ell$-th UT is given by
\begin{equation}
y_{\ell,i} = x_{\ell,i} + \gamma x_{\ell-1,i} + z_{\ell,i},
\end{equation}
where $\gamma \in [0,1]$ represents the inter-cell interference level and $z_{\ell,i} \sim \Cc\Nc(0,1)$.  The extension of results in previous sections to the (complex-valued) Soft-Handoff model is easy  and done in the usual way \cite{songnam2011DAS}.  In this example, thanks to the dual-diagonal structure of the channel matrix, the system matrix is guaranteed to have rank $L$.
In Fig.~\ref{simulation_downlink}, we compare various coding strategies where the upper bound and achievable rates for CDPC
are provided by \cite{simeone2009downlink}. It is remarkable that RCoF can achieve the upper bound when $R_{0} \leq 4$ bits and
outperforms the other schemes up to $R_{0} \approx 6$ bits.
Notice that when $\gamma=1$ (e.g., integer channel matrix), RCoF almost achieves the upper bound, showing better performance than other schemes.
Also, from the Fig.~\ref{simulation_DL}, we can see that RCoF is a good scheme when $R_{0}$ is small and SNR is high, i.e., small cell networks
with finite-backhaul capacity. Not surprisingly, RQCoF approaches the performance of RCoF
within the shaping loss of $\approx 0.25$ bits/symbol, as already noticed in the uplink case \cite{songnam2011compute}.

For the RQCoF, there would be a concern on rank-deficiency of system matrices $\Qm$ in particular when $p$ is small,
since every UT selects its own linear combination coefficients independently of the other nodes.
This problem can be avoid by scheduling since it can select a group of UTs (or ATs) for which the system matrix is invertible.
In fact, this is a complex combinatorial optimization problem, which in some cases, can be formulated as the maximization of linear function
over matroid constraint \cite{edmonds1971matroids} and thus, greedy algorithm yields provably good performance.
The following is the example that greedy algorithm is optimal.
Consider a DAS system with  $K$ UTs and $L$ ATs where $K \geq L$ and we consider the user selection that finds the subset
of UTs to maximize the symmetric rate subject to full-rank constraint of system matrix.
Independently of the user selection algorithm, we can obtain the coefficients of the linear combination of the $k$-th UT
(e.g.,  $k$-th row of $\Qm$) and  the variance of effective noise (i.e., $\sigma_{\varepsilon_{k}}^2$ for $k=1,\ldots,K$) that determines
the achievable rate, for the given $\Hm \in \RR^{K \times L}$.
Let $\Kc$ be the subset of row indices $[1:K]$.
Also, let $\Qm(\Kc)$ denote the submatrix of $\Qm$ consisting of $k$-th rows for $k \in \Kc$.
Assuming that $\Qm$ has rank $L$, the user selection problem of finding $L$ UTs  can be formulated as
\begin{eqnarray}
\underset{\Kc \subset [1:K]}{\operatornamewithlimits{arg\,min } } && \max\{\sigma_{\varepsilon_{k}}^2: k \in \Kc\}\label{eq:object}\\
\hbox{subject to}&& \hbox{Rank}_{p}(\Qm(\Kc)) = L\label{constraint}
\end{eqnarray}  We first give the definition of matroid and subsequently, show that the above problem  is equivalent to the maximization of linear function over matroid constraint. Matroids are structures that generalize the concept of linear independence for general sets. Formally, we have \cite{edmonds1971matroids}:
\begin{definition}\label{def:matroid} A matroid $\Mc$ is a tuple $\Mc=(\Omega,\Ic)$,
where $\Omega$ is a finite ground set and $\Ic \subseteq 2^{\Omega}$ (the power set of $\Omega$)
is a collection of independent sets, such that:
\begin{enumerate}
\item $\Ic$ is nonempty, in particular, $\phi \in \Ic$
\item $\Ic$ is downward closed; i.e., if $\Yc \in \Ic$ and $\Xc \subseteq \Yc$, then $\Xc \in \Ic$
\item if $\Xc,\Yc \in \Ic$, and $|\Xc| < |\Yc|$, then $\exists y \in \Yc\setminus \Xc$ such that $\Xc \cup \{y\} \in \Ic$.
\end{enumerate}
\end{definition}
Let $\Omega=[1:K]$ and $\Ic =\{\Kc \subset [1:K] :\Qm(\Kc) \hbox{ has linearly independent rows}\}$.
From Definition \ref{def:matroid}, $\Mc=(\Omega,\Ic)$ forms a so-called linear matroid. Then, the optimization problem (\ref{eq:object})-(\ref{constraint}) is equivalent to
\begin{eqnarray}
\underset{\Kc \subset [1:K]}{\operatornamewithlimits{arg\,max } } && \sum_{k \in \Kc} 1/\sigma_{\varepsilon_{k}}^2\\
\hbox{subject to}&& \Kc \in \Ic
\end{eqnarray}   This can be easily proved by the fact that $\Qm$ has rank $L$ and constraint is matroid.  Rado and Edmonds proved that the Best-In-Greedy algorithm (See Algorithm $1$) finds an optimal solution  \cite{edmonds1971matroids}.   Detailed scheduling algorithms for various scenarios are omitted because of space limitation (See \cite{songnam2011DAS}).

\begin{algorithm}
\caption{Best-In-Greedy Algorithm}
\begin{description}
\item[Input: ] $\Mc=(\Omega,\Ic)$ and $w_{k}=1/\sigma_{\varepsilon_{k}}^2$ for $k \in [1:K]$
\item[step 0. ] Sort $[1:K]$ such that $w_{1} \geq w_{2} \geq \cdots \geq w_{K}$
\item[] Initially $k=1$ and $\Kc = \phi$
\item[step 1. ] If $\mbox{Rank}_{p}(\Qm(\Kc \cup \{k\})) > \mbox{Rank}_{p}(\Qm(\Kc))$, then $\Kc \leftarrow \Kc \cup \{k\}$
\item[step 2. ] Set $k=k+1$
\item[step 3. ] Repeat until $\mbox{Rank}_{p}(\Qm(\Kc )) = L$
\end{description}
\end{algorithm}

\begin{figure}
\centerline{\includegraphics[width=9cm]{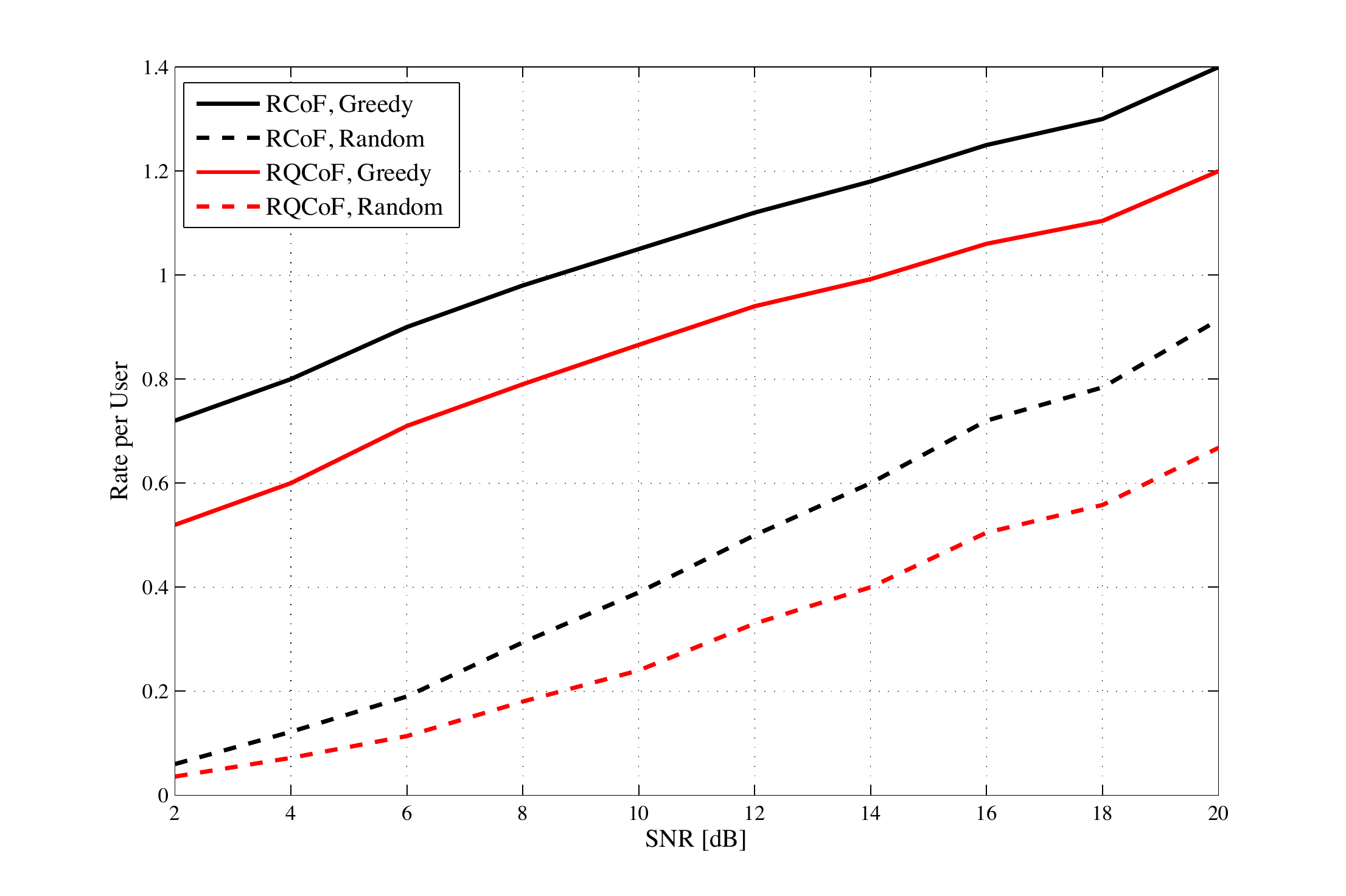}}
\caption{Achievable rates per user as a function of SNRs, for finite capacity $R_{0}=3$ bits and $p=17$ for RQCoF.}
\label{simulation_scheduling}
\end{figure}

In Fig.~\ref{simulation_scheduling}, we consider a DAS with channel matrix $\Hm \in \RR^{20 \times 5}$,  with i.i.d. Gaussian distributed elements $\sim \Nc(0,1)$.  In our simulation we assumed that if the resulting system matrix after greedy selection is rank deficient then the achieved symmetric rate of all users is zero, for that specific channel realization. Then, we computed the average achievable rate with user selection, by Monte Carlo averaging with respect to the  random channel matrix.
Random selection indicates that $5$ UTs are randomly and uniformly chosen out of the $20$ UTs.
As shown in Fig.~\ref{simulation_scheduling}, RCoF has the rank-deficiency when using random selection, although the rank of the
resulting $5 \times 5$ matrix  over $\RR$ is equal to 5 with probability 1.
However, it is remarkable that RQCoF with greedy user selection does not suffer from the rank-deficiency problem, even for
relatively small values of $p$ (e.g., $p=17$). This is indicated by the fact that the gap from the RCoF is essentially equal to the
the shaping loss, as in the case where the full-rank system matrix is guaranteed by assumption.

\end{document}

%% file: macros.tex
\long\def\comment#1{}

\newfont{\bbb}{msbm10 scaled 700}

\newfont{\bb}{msbm10 scaled 1100}

\newcommand{\RR}{\mbox{\bb R}}

\newcommand{\ZZ}{\mbox{\bb Z}}

\newcommand{\EE}{\mbox{\bb E}}

\newcommand{\av}{{\bf a}}

\newcommand{\cv}{{\bf c}}
\newcommand{\dv}{{\bf d}}

\newcommand{\hv}{{\bf h}}

\newcommand{\qv}{{\bf q}}

\newcommand{\uv}{{\bf u}}
\newcommand{\wv}{{\bf w}}

\newcommand{\xv}{{\bf x}}

\newcommand{\zv}{{\bf z}}
\newcommand{\zerov}{{\bf 0}}

\newcommand{\Am}{{\bf A}}

\newcommand{\Fm}{{\bf F}}
\newcommand{\Gm}{{\bf G}}
\newcommand{\Hm}{{\bf H}}
\newcommand{\Id}{{\bf I}}

\newcommand{\Qm}{{\bf Q}}

\newcommand{\Um}{{\bf U}}
\newcommand{\Wm}{{\bf W}}

\newcommand{\Cc}{{\cal C}}

\newcommand{\Ic}{{\cal I}}

\newcommand{\Kc}{{\cal K}}

\newcommand{\Mc}{{\cal M}}
\newcommand{\Nc}{{\cal N}}

\newcommand{\Sc}{{\cal S}}

\newcommand{\Vc}{{\cal V}}
\newcommand{\Xc}{{\cal X}}
\newcommand{\Yc}{{\cal Y}}

\newcommand{\lambdav}{\hbox{\boldmath$\lambda$}}

\newcommand{\muv}{\hbox{\boldmath$\mu$}}

\newcommand{\SNR}{{\sf SNR}}

\newcommand{\transp}{{\sf T}}
